\documentclass[]{spie}  

\newcommand{\bicep}{B\textsc{icep}}

\newcommand{\ukrts}{$\mu$K$\sqrt{\textrm{s}}$}

\newcommand{\lcdm}{$\Lambda$CDM }

\usepackage{amsmath,amsfonts,amssymb}
\usepackage{graphicx}
\usepackage[colorlinks=true, allcolors=blue]{hyperref}
\usepackage{siunitx}
\usepackage{booktabs}
\usepackage{caption} 
\usepackage{color}
\usepackage{tablefootnote}

\title{2017 upgrade and performance of BICEP3: a 95GHz refracting telescope for degree-scale CMB polarization}

\author[a]{Jae Hwan Kang}
\author[b]{P.~A.~R.~Ade}
\author[c]{Z.~Ahmed}
\author[d]{R.~W.~Aikin}
\author[e]{K.~D.~Alexander}
\author[e]{D.~Barkats}
\author[f]{S.~J.~Benton}
\author[g]{C.~A.~Bischoff}
\author[d,h]{J.~J.~Bock}
\author[e]{H.~Boenish}
\author[e]{R.~Bowens-Rubin}
\author[d]{J.~A.~Brevik}
\author[e]{I.~Buder}
\author[i]{E.~Bullock}
\author[e,j]{V.~Buza}
\author[e]{J.~Connors}
\author[e]{J.~Cornelison}
\author[h]{B.~P.~Crill}
\author[n]{M.~Crumrine}
\author[e]{M.~Dierickx}
\author[k]{L.~Duband}
\author[j]{C.~Dvorkin}
\author[l,m]{J.~P.~Filippini}
\author[n]{S.~Fliescher}
\author[a]{J.~A.~Grayson}
\author[n]{G.~Hall}
\author[o]{M.~Halpern}
\author[e]{S.~Harrison}
\author[d,h]{S.~R.~Hildebrandt}
\author[p]{G.~C.~Hilton}
\author[d]{H.~Hui}
\author[a,c,p]{K.~D.~Irwin}
\author[e,q]{K.~S.~Karkare}
\author[a]{E.~Karpel}
\author[r]{J.~P.~Kaufman}
\author[r]{B.~G.~Keating}
\author[d]{S.~Kefeli}
\author[a]{S.~A.~Kernasovskiy}
\author[e,j]{J.~M.~Kovac}
\author[a,c]{C.~L.~Kuo}
\author[q]{N.~A.~Larsen}
\author[n]{K.~Lau}
\author[q]{E.~M.~Leitch}
\author[d]{M.~Lueker}
\author[h]{K.~G.~Megerian}
\author[d]{L.~Moncelsi}
\author[s]{T.~Namikawa}
\author[f,t]{B.~Netterfield}
\author[h]{H.~T.~Nguyen}
\author[d,h]{R.~O'Brient}
\author[a,c]{R.~W.~Ogburn~IV}
\author[g]{S.~Palladino}
\author[i,n]{C.~Pryke}
\author[e]{B.~Racine}
\author[e]{S.~Richter}
\author[d]{A.~Schillaci}
\author[n]{R.~Schwarz}
\author[q,u]{C.~D.~Sheehy}
\author[d]{A.~Soliman}
\author[e]{T.~St.~Germaine}
\author[d,h]{Z.~K.~Staniszewski}
\author[d]{B.~Steinbach}
\author[b]{R.~V.~Sudiwala}
\author[d,r]{G.~P.~Teply}
\author[a,c]{K.~L.~Thompson}
\author[a]{J.~E.~Tolan}
\author[b]{C.~Tucker}
\author[h]{A.~D.~Turner}
\author[g]{C.~Umilt\`{a}}
\author[q,v]{A.~G.~Vieregg}
\author[d]{A.~Wandui}
\author[h]{A.~C.~Weber}
\author[o]{D.~V.~Wiebe}
\author[n]{J.~Willmert}
\author[e,j]{C.~L.~Wong}
\author[a,q]{W.~L.~K.~Wu}
\author[a]{H.~Yang}
\author[a,c]{K.~W.~Yoon}
\author[d]{C.~Zhang}

\affil[a]{Department of Physics, Stanford University, Stanford, CA 94305, USA}
\affil[b]{School of Physics and Astronomy, Cardiff University, Cardiff, CF24 3AA, United Kingdom}
\affil[c]{Kavli Institute for Particle Astrophysics and Cosmology, SLAC National Accelerator Laboratory,Menlo Park, CA 94025, USA}
\affil[d]{Department of Physics, California Institute of Technology, Pasadena, CA 91125, USA}
\affil[e]{Harvard-Smithsonian Center for Astrophysics, Cambridge, MA 02138, USA}
\affil[f]{Department of Physics, University of Toronto, Toronto, Ontario, M5S 1A7, Canada}
\affil[g]{Department of Physics, University of Cincinnati, Cincinnati, OH 45221, USA}
\affil[h]{Jet Propulsion Laboratory, Pasadena, CA 91109, USA}
\affil[i]{Minnesota Institute for Astrophysics, University of Minnesota, Minneapolis, MN 55455, USA}
\affil[j]{Department of Physics, Harvard University, Cambridge, MA 02138, USA}
\affil[k]{Service des Basses Temp´eratures, Commissariat `a l’Energie Atomique, 38054 Grenoble, France}
\affil[l]{Department of Physics, University of Illinois at Urbana-Champaign, Urbana, IL 61801, USA}
\affil[m]{Department of Astronomy, University of Illinois at Urbana-Champaign, Urbana, IL 61801, USA}
\affil[n]{School of Physics and Astronomy, University of Minnesota, Minneapolis, MN 55455, USA}
\affil[o]{Department of Physics and Astronomy, University of British Columbia,Vancouver, British Columbia, V6T 1Z1, Canada}
\affil[p]{National Institute of Standards and Technology, Boulder, CO 80305, USA}
\affil[q]{Kavli Institute for Cosmological Physics, University of Chicago, Chicago, IL 60637, USA}
\affil[r]{Department of Physics, University of California at San Diego, La Jolla, CA 92093, USA}
\affil[s]{Leung Center for Cosmology and Particle Astrophysics, National Taiwan University, Taipei 10617, Taiwan}
\affil[t]{Canadian Institute for Advanced Research, Toronto, Ontario, M5G 1Z8, Canada}
\affil[u]{Physics Department, Brookhaven National Laboratory, Upton, NY 11973}
\affil[v]{Department of Physics, Enrico Fermi Institute, University of Chicago, Chicago, IL 60637}

\authorinfo{Corresponding author: J. Kang, jaehwan@stanford.edu}

\begin{document} 
	\maketitle
	
	\begin{abstract}
		\bicep3 is a 520mm aperture on-axis refracting telescope observing the polarization of the cosmic microwave background (CMB) at 95GHz in search of the B-mode signal originating from inflationary gravitational waves. \bicep3's focal plane is populated with modularized tiles of antenna-coupled transition edge sensor (TES) bolometers. \bicep3 was deployed to the South Pole during 2014-15 austral summer and has been operational since. During the 2016-17 austral summer, we implemented changes to optical elements that lead to better noise performance. We discuss this upgrade and show the performance of \bicep3 at its full mapping speed from the 2017 and 2018 observing seasons. \bicep3 achieves an order-of-magnitude improvement in mapping speed compared to a Keck 95GHz receiver. We demonstrate 6.6\ukrts\, noise performance of the \bicep3 receiver.
	\end{abstract}
	
	
	\section{INTRODUCTION}
	\label{sec:intro}  
	The cosmic microwave background (CMB) has been observed extensively to probe the early history and evolution of the Universe, bolstering the standard \lcdm cosmology. However, the standard \lcdm cosmology fails to provide an explanation for the homogeneity, isotropy, and flatness of the observable Universe. Inflation is the leading framework that resolves these problems by assuming an epoch of exponential expansion prior to the standard Big Bang expansion. Inflation stretches the tensor perturbations of the metric and produces a stochastic gravitational wave background. Since the inflationary gravitational wave (IGW) background is the only source known to produce the curly pattern in polarization, so-called `B-mode' polarization, on the CMB at the last scattering surface, the detection of B-modes at degree angular scales is a valid test for this framework\cite{Kamionkowski2016}.
    The level of the B-mode power from IGW is related to the energy scale of the inflation, and it is characterized by the ratio of the amplitude of the tensor perturbation to the amplitude of the scalar perturbation, called the tensor-to-scalar ratio `$r$'. The predicted level of the B-mode power is very small and requires deep CMB polarization maps.
    
	The \bicep/Keck Collaboration has deployed a series of telescopes to the South Pole in search of the B-mode signal potentially originating from the IGW. With these small aperture telescopes, we focus on scanning a small patch of the sky to obtain deep polarization maps at degree angular scales. Since the detection of the B-mode at degree angular scales by \bicep2 which had operated at 150GHz\cite{BKI}, we have continued to observe the same patch at multiple frequency bands to separate CMB polarization from foreground components that can also contribute to the B-mode power. Keck Array has been equipped with five \bicep2-size telescopes at different frequencies and operational since 2012. Utilizing the capacity to house five receivers on the Keck mount, we have deployed and replaced receivers at frequencies of 95, 150, 220, and 270GHz. \bicep3 is the third generation of the family with a larger aperture and faster optics, hosting more detectors on its focal plane than \bicep2 or Keck Array receivers. With 2400 optically coupled transition edge sensor (TES) bolometers at its full capacity, \bicep3 was designed to achieve ten times the throughput of a single Keck 95GHz receiver. The design of \bicep3 was discussed in detail in Ref.~\citenum{Ahmed14}.
	
	The initial performance from the engineering season in 2015 was presented in Ref.~\citenum{wu15}. There were significant upgrades and fixes based on what we investigated from the engineering season, and the preliminary performance from the first science season in 2016 was presented in Refs.~\citenum{Grayson16},\citenum{Hui16},\citenum{Karkare16} including further details of \bicep3 design and operation schemes. The first science season still exhibited an elevated level of per-detector noise compared to Keck 95GHz receivers. This proceeding discusses the upgrades and fixes we implemented for the second science season in 2017 and presents \bicep3 performing at its full stable capacity.

	\section{INSTRUMENT UPGRADE}
	\label{sec:instument}
	
	A brief instrument overview and discussion of the upgrades we implemented during the 2016-17 austral summer.

	\begin{figure} [t]
		\begin{center}
			\begin{tabular}{c} 
				\includegraphics[width=.82\textwidth]{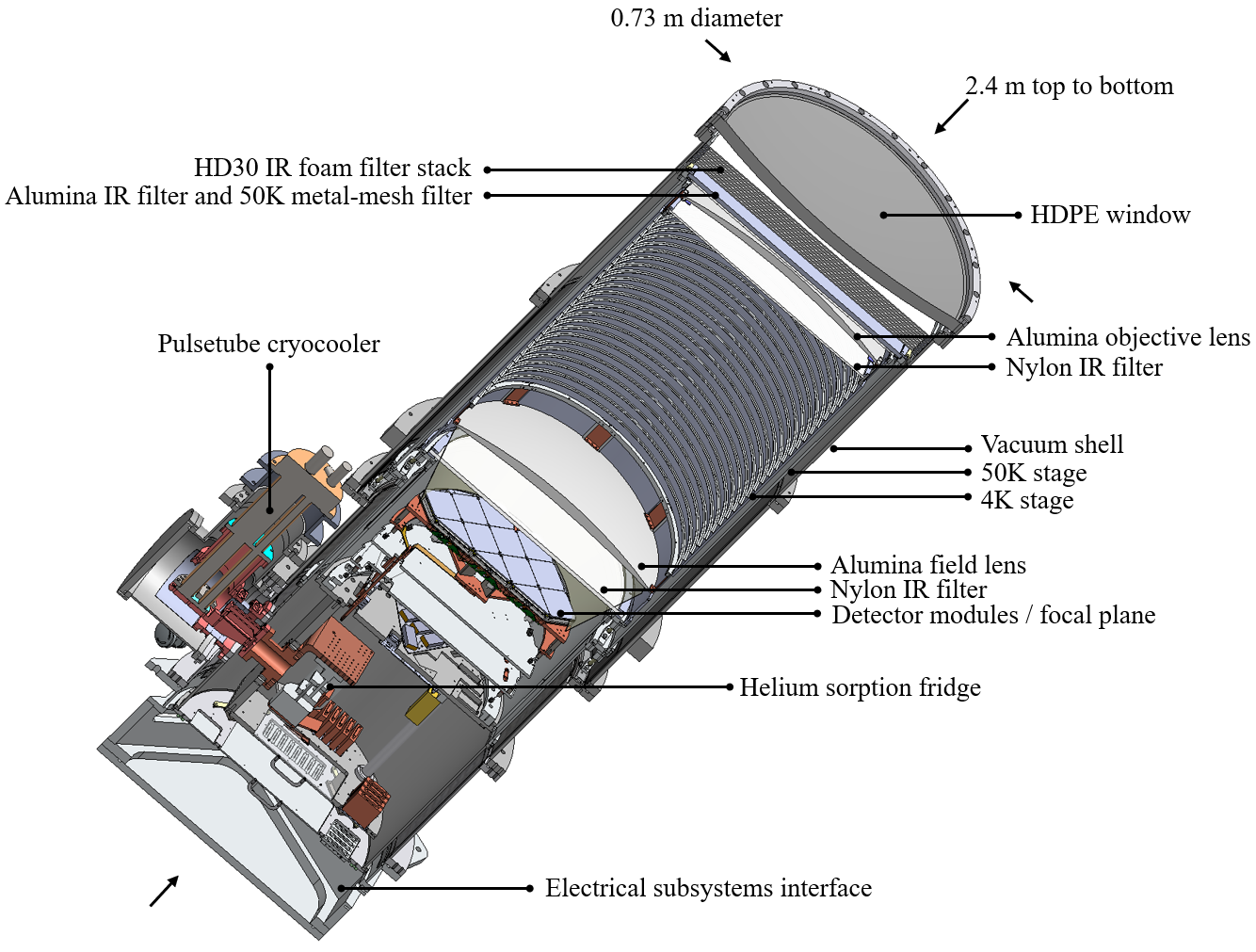}				
			\end{tabular}
		\end{center}
		\caption[example] 
		{ \label{fig:B3cutaway} Cutaway view of the \bicep3 receiver. Update from Ref.~\citenum{Grayson16} is the replacement of the  metal-mesh IR filter stack with the  HD30 IR foam filter stack.}
	\end{figure} 
	
	\subsection{Instrument Overview}
	Small aperture designs provide effective systematics control while achieving deep sensitivity at degree angular scales where the B-mode power spectrum from IGW peaks. Figure~\ref{fig:B3cutaway} shows a cutaway view of the \bicep3 receiver computer-aided design model. The receiver consists of three concentric cylinders at different temperatures, radiatively insulated from one another and supported by low thermal conductivity materials. The outermost shell at room temperature encloses a vacuum. The top of the vacuum shell is capped by a high-density polyethylene (HDPE) window. A PT415 Pulsetube cryocooler\footnote{Cryomech Inc., Syracuse, NY 13211, USA (www.cryomech.com)} cools the inner shells to 50K and 4K stages respectively through oxygen-free high-conductivity (OFHC) copper heat straps. The top half of the 4K stage contains the objective and field lenses. The bottom half of the 4K stage contains the sub-Kelvin structure which houses the focal plane unit (FPU). The sub-Kelvin structure is cooled by a three-stage helium sorption fridge\footnote{Chase Research Cryogenics Ltd., Sheeld, S10 5DL, UK (www.chasecryogenics.com)} to reach 2K, 350mK, and 250mK stages. The FPU is on the 250mK stage and populated with 20 modularized detector tiles. Temperature control modules (TCMs) provide stable thermal control of FPU to keep temperature fluctuation small around set temperature points. The pulsetube cryocooler provides continuous cooling, but the helium sorption fridge requires a recycling period to condense helium in the still.
    
	\bicep3's aperture doubled the existing \bicep2/Keck receivers. As a result, the loading through the window ($\sim$110W) would exceed the cooling power of the cryocooler. It is therefore imperative that the first stage of IR filtering is substantially reflective (or scattering) rather than absorptive, therefore reducing the thermal loading on the cryocooler. Our IR filter sequence starts just below the 300K window and continues all the way down to 4K to reduce loading.
    
    Just above the detector modules are metal mesh low pass edge filters\cite{Ade06} which reject above-band radiative loading and limit high-frequency photons from directly coupling to the TES bolometers. The detectors, first stage SQUID amplifiers and multiplexing electronics are packaged in the detector module\cite{Hui16}. The final detector signal is amplified by SQUID series arrays (SSAs) at the 4K stage. The bias settings and readout of the detector channels are controlled by Multi-Channel-Electronics (MCEs) developed by the University of British Columbia which are attached to the bottom interface of the receiver at room temperature.
	
	\begin{figure} [t]
		\begin{center}
			\begin{tabular}{c} 
				\includegraphics[width=.4\textwidth]{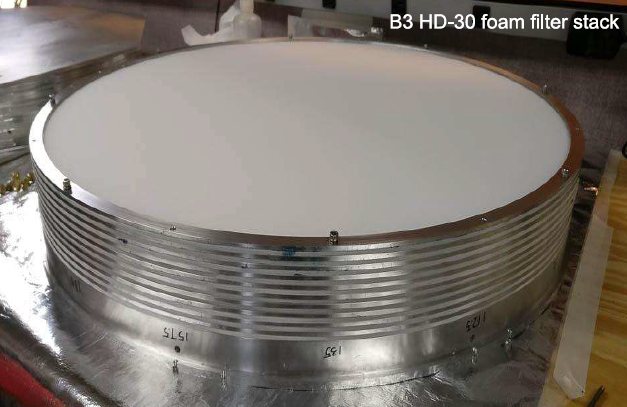}
			\end{tabular}
		\end{center}
		\caption[example] 
		{ \label{fig:B3HD30} The HD30 IR foam filter stack.}
	\end{figure} 
	
	\subsection{The IR Filter Stack Replacement}
	\label{sec:IRfilters}
	\begin{table}
	\centering
	\caption[Loading on each temperature stages
	]{
	Loading on each temperature stages
	}
	\label{tab:IRload}
		\begin{tabular}{l l l} 
		\toprule 
		Stages   &    Power in 2016 [W] & Power in 2017 [W]\\
		\toprule 
		Window           &  $\sim$110 & $\sim$110 \\
		50~K optics tube &  19        & 12 \\
		4~K optics tube  &  0.18      & 0.15 \\
		350~mK stage     &  $9.05\times 10^{-5}$     & $8.4\times10^{-5}$ \\
		Focal Plane (250~mK) &  $3.5\times10^{-6}$     & $3.35\times10^{-6}$ \\
		\bottomrule
		\vspace{1mm}
		\end{tabular}
	\end{table}
			
	\begin{table}
	\centering
	\caption[In-band optical load in \bicep3]{Per-detector in-band optical load during the 2016 and 2017 seasons, quantified as both incident power and Raleigh-Jeans temperature. Season year is specified for the IR filter stack.}
	\label{tab:preddetload}
		\begin{tabular}{l c c c c} 
		\toprule 
		Source   &    Load (2016) [pW] & Load (2017) [pW] & $T_\mathrm{RJ}$ (2016) [K] & $T_\mathrm{RJ}$ (2017) [K] \\
		\toprule 
		4K lenses \& elements & 0.15 & 0.15 & 1.0 & 1.0 \\
		50K alumina filter   & 0.10  & 0.10 & 0.80 & 0.80\\
		Metal-mesh filters (2016) & 0.63&- & 5.2  & - \\
		HD-30 foam filters (2017) & - & 0.10 & - & 0.79\\
		Window     &  0.69   & 0.72 & 5.9 & 5.9\\
		\midrule
		Total cryostat internal &1.57 &1.07 &12.9 &8.5 \\			\midrule
		Forebaffle   &  0.308 & 0.136 & 2.67 & 1.14 \\
		Atmosphere   &  1.13  & 1.18 & 9.91 & 9.91 \\
		CMB          &  0.121 & 0.126 & 1.11 & 1.11 \\
		\midrule
		Total    &  3.13 & 2.51 & 26.6 & 20.7 \\
		\bottomrule
		\vspace{1mm}
		\end{tabular}
	\end{table}

	We had placed the stack of 10 metal-mesh IR-reflective filters beneath the window to reduce the load on the pulsetube cryocooler. This filter stack effectively reduced the loading down to 19W, well within the capacity of the pulsetube (Table~\ref{tab:IRload}). In the austral summer of 2016-17, we replaced this capacitive metal-mesh IR filter stack with a high-density polyethylene (HD30\footnote{HD30: High-density polyethylene, $30kg/m^3$. http://www.zotefoams.com/}) foam filter stack. Each foam filter is 1/8 inch thick, each spaced 1/8 inch apart. Figure~\ref{fig:B3HD30} shows the foam filter stack. This change further reduced the IR loading while increasing the in-band transmission.  This thin foam IR filter scheme was investigated by Choi, et al\cite{choi13}, although they used a different type of foam. The bottom layer of the absorptive foam filters emits thermal radiation at much lower temperature than the top layer. The low index of refraction of the foam reduces reflection and makes anti-reflection coating unnecessary.
	
    Table~\ref{tab:IRload} shows the loading on each temperature stages. The power on the 50K stage was reduced by 7W in 2017, clearly showing the effectiveness of foam filters. Consequently the temperature of the 50K stage was reduced, which further reduces the loading on subsequent thermal stages. Table~\ref{tab:preddetload} shows per-detector in-band optical load, quantified as both incident power and Raleigh-Jeans temperature. The source is listed in time reverse sense. Reduced response from the forebaffle indicates less wide-angle scattering from the absorptive foam filters than the reflective metal-mesh filters. The in-band transmission through the stack of 10 filters increased from $\sim$92\% via mesh filters to $\sim$100\% via foam filters in lab tests. Overall, we have higher optical efficiency (see Section~\ref{sec:OptEff}) and let in more CMB signal with less instrumental loading.
 
 Without a long history of lab testing, including repeated
vacuum cycling, it is worth considering the long-term 
performance of the HD-30 filters, both for IR blocking and 
for minimal effects on microwave transmission.  We note that the Keck Array receivers all have had HD-30 windows exposed
to vacuum on one side for long periods, with no apparent
degradation of receiver performance.  However, we note that HD-30
entirely within a vacuum will outgas the N2 gas from 
its cells, then distort when brought back to 1 atmosphere.
The 1/8 inch sheets we use for filters do return to nearly
their original state at 1 atm, but will pull away from 
their mounting rings unless firmly attached, and also have 
a residual bow that does not relax over time. Our current  
strategy is to replace the filter stack between vacuum 
cycles. We will also do further testing, including the
investigation of LD-24 (based on low density polyethylene), 
which does not show the same residual distortion after an 
initial recovering stage at 1 atm and seems to have similar 
IR-blocking efficiency. 
	
	\subsection{Module and Low Pass Edge Filter Replacement}
	We estimated the noise equivalent temperature (NET) of each detector module in the 2016 season. During the 2016-17 austral summer, we replaced the four worst performing modules. The modularized detector tile design of the \bicep3 focal plane allowed us to conveniently replace individual tiles to improve overall receiver performance. In some modules that we retained, wirebonds that failed when the receiver got cold were fixed on-site.

	\begin{figure} [t]
	\begin{center}
		\begin{tabular}{c}
			\includegraphics[width=0.5\textwidth]{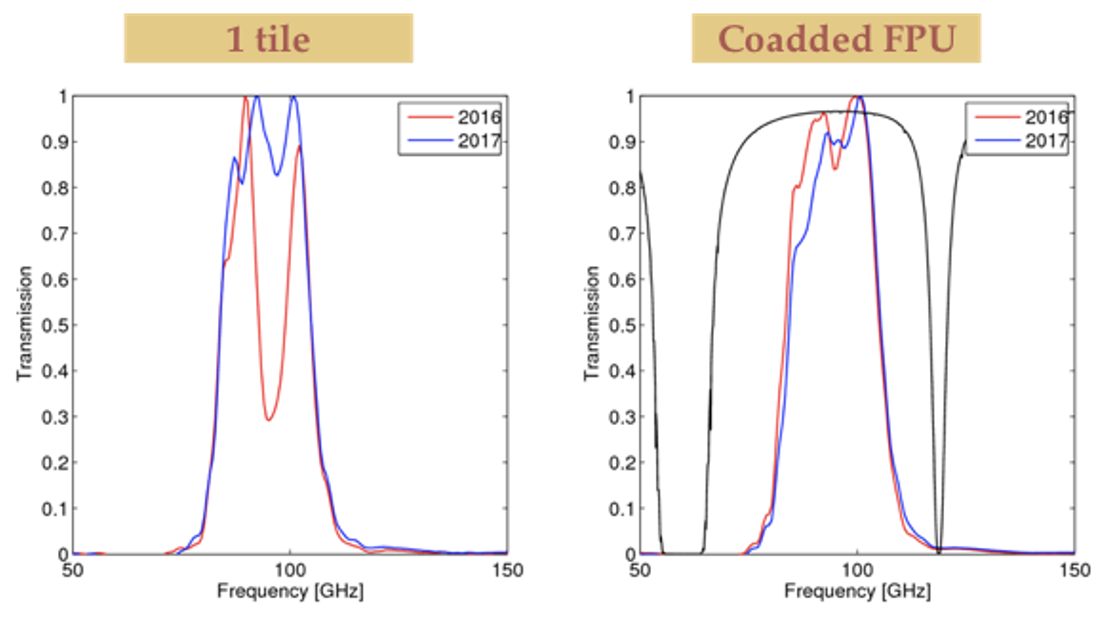}
		\end{tabular}
	\end{center}
	\caption[example] 
	{ \label{fig:Bandpass} A tile-averaged (\textit{left}) and the array-averaged (\textit{right}) frequency response spectrum from the 2016 season (\textit{red}) and the 2017 season (\textit{blue}). Each bandpass is normalized to one at its peak. The black solid line indicates the atmosphere transmission.}
	\end{figure} 

	  Fourier Transform Spectrometer (FTS) measurements of the bandpasses exhibited anomalous features in most tiles in 2016. When the \bicep3 receiver was opened to the focal plane for module replacement at the end of 2016 season, we found that the edge filters had delaminated. The delaminated edge filters apparently caused the anomalous bandpasses and the wide distribution of optical efficiencies. For the 2017 season we replaced the filters. Some were from a new fabrication and some had been manufactured several years before (for Keck Array), and the anomaly disappeared (Figure~\ref{fig:Bandpass}). Despite the anomaly in individual detectors in the 2016 season, the overall receiver spectral performance passes initial checks, but detailed systematics study associated with these bandpass profiles is in progress.
	
	\section{OBSERVING STRATEGY}
	\begin{figure} [t]
		\begin{center}
			\begin{tabular}{c}
				\includegraphics[width=0.95\textwidth]{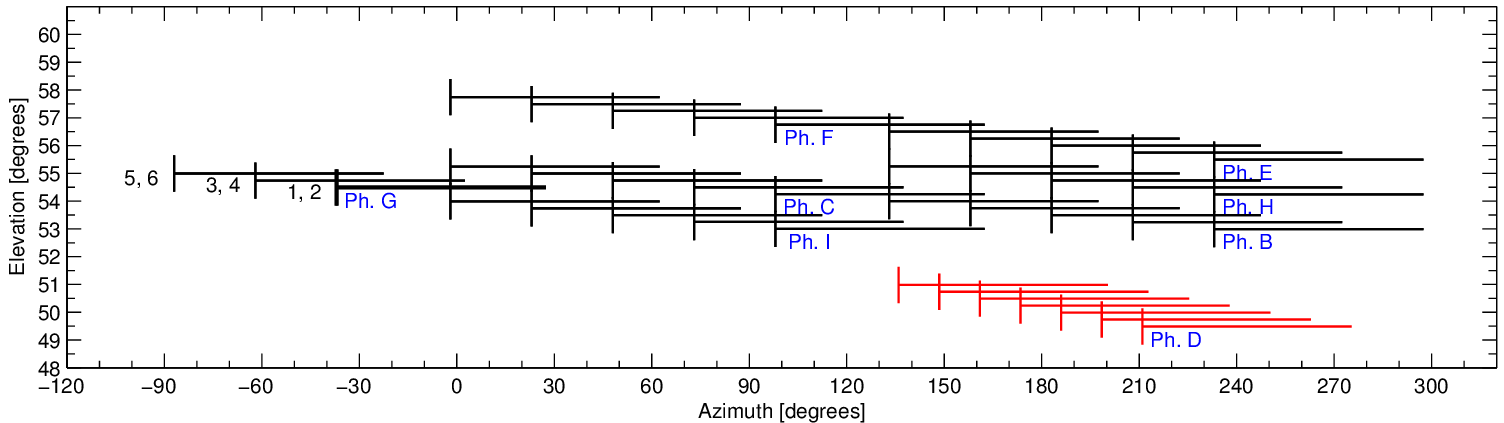}
			\end{tabular}
		\end{center}
		\caption[example]
		{ \label{fig:pat} Observing pattern of a typical three-day schedule in ground-based coordinates. The first scanset of phase G is shown in bold. Horizontal lines indicate the field scans and the vertical lines indicate the bracketing elevation nods. The telescope scans at a fixed elevation during each scanset. For the CMB field scans, we observe two scansets before changing elevation. Phase D is on the galactic plane.}
	\end{figure} 
	
	\bicep3 uses the same mount in the Dark Sector Laboratory (DSL) of the South Pole Station that also supported previous \bicep\, experiments. We observe the same part of the sky as the previous \bicep\, experiments did and Keck Array currently does, spanning $-60^\circ< \rm{RA} <60^\circ$ and $-70^\circ<\delta<-40^\circ$. With its larger field of view, \bicep3 covers more area, spanning an effective sky area of about $600\rm{deg}^2$, larger than the $400\deg^2$ of \bicep2/Keck. At the South Pole, the sky moves along the azimuth but does not rise or set in elevation. Thus we can track the same patch at any time while detectors are working. 
    
	Figure~\ref{fig:pat} shows the observing pattern of a typical 3-day schedule in ground-based coordinates. The fundamental element of observation is a `scanset' which consists of 50 back-and-forth scans at a fixed elevation at a speed of $2.8\deg /s$, shown as a horizontal line in the figure. At the start and end of a scanset are bracketing elevation nods to calibrate and monitor detector gains in airmass units, appearing as a vertical line in the figure. Each scanset lasts for about 50 minutes and the sky drifts by $12.5^\circ$ during that time. For CMB field scans, we observe two scansets before changing the elevation offset and azimuth center, so the next pair of scansets starts at the shifted azimuth center by $25^\circ$. The azimuth center is set so that the average RA over a pair of scansets at a fixed elevation is at RA=0. Phases D consists of seven scansets on the galactic plane. We run schedules with different boresight deck angles and different combination of elevation offsets of the phases for systematics control and uniform coverage.
	
	The duration of the schedule is determined by the fridge hold time from the fridge cycle. We ran observation schedules with a 2-day cadence in 2016. In 2017, the upgraded optics reduced the IR loading which improved He condensation by lowering 4K temperatures. We tweaked the set temperature points of the TCMs on the focal plane structure, further reducing the load on the sorption fridge. We tested the fridge cycle to achieve near 100 hours of fridge hold time and ran a few 4-day cadence schedules. We chose to run 3-day observing schedules to gain the margin needed to insure reliable control of the focal plane temperature and a consistent observation environment. Given the same 6 hours of fridge cycling time, a 3-day schedule provides 4\% more available observing time than a 2-day schedule. By devoting the third day to CMB observation instead of including galaxy observation, we can spend 10\% more time on CMB than the previous schedule. Observing dates for each season can vary a bit for running tests and maintenance operations. Overall, we spent 8.5\% more time observing the CMB in 2017 (5043 scansets) than in 2016 (4647 scansets). In addition to the improved detector performance (Section~\ref{sec:performance}), the extended schedules help collect more data per year.
	
	\section{DETECTOR PERFORMANCE}
	\label{sec:performance}
	\bicep3's initial performance from the 2015 engineering season is reported in Ref.~\citenum{wu15}. After the upgrade to a full complement of 20 tiles of detectors, the performance in the second season (as the first science season) is summarized in Ref.~\citenum{Grayson16}. What follows is the update to the performance from the 2016 season with full-season data. We compare the results between the full-season data from both 2016 and 2017 observing seasons and include the preliminary results from the current 2018 observing season.

	\subsection{Detector Yield}
	The aforementioned upgrades improved overall detector yield from 79\% to 84\% from 2016 to 2017. We had 1006 detector pairs working in 2017. After choosing the optimal TES bias points, which are set for detectors sharing the same MUX column, we have 856 performing pairs in 2017, an improvement from 763 pairs in 2016.

	\subsection{Optical Efficiency}
    \label{sec:OptEff}
	\begin{figure} [t]
	\begin{center}
		\begin{tabular}{c}
			\includegraphics[width=0.4\textwidth]{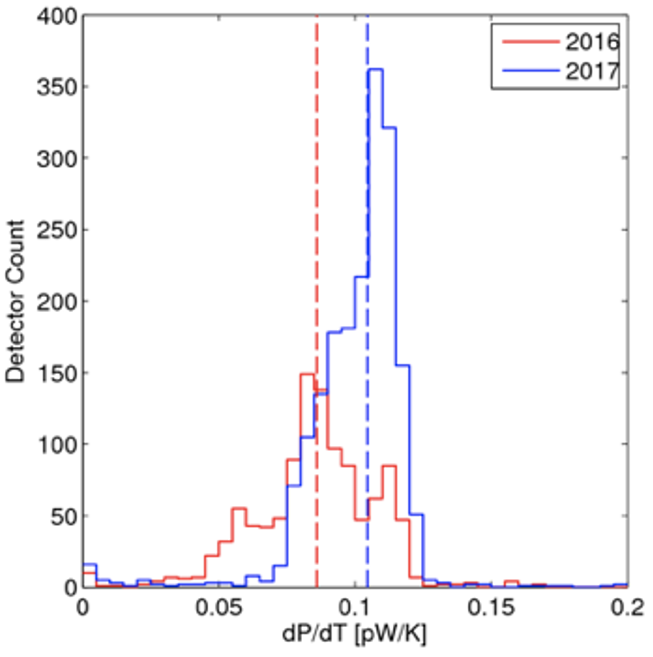}
		\end{tabular}
	\end{center}
	\caption[example] 
	{ \label{fig:OptEff} Optical efficiency from 2016 season (red) and 2017 season (blue). The median values are marked by vertical dotted lines. This measurement is done in aluminum transition, which has a lower yield than science observing setting.}
	\end{figure}
	We placed a beam filling blackbody source (a flat Eccosorb\textsuperscript{\textcircled{R}} sheet) on top of the window in a container to hold liquid nitrogen. Measuring the detector load curves under the source at ambient temperature and at the temperature of liquid nitrogen, we can determine the optical efficiency of the detectors and the results are shown in Figure~\ref{fig:OptEff}. The median optical efficiency is improved from $0.08\rm{pW}/\rm{K_{RJ}}$ to $0.1\rm{pW}/\rm{K_{RJ}}$, which corresponds to the efficiency rising from $\sim$24\% to $\sim$29\%. The distribution is also tighter in 2017 season.

	\subsection{Timestream-based NET}
	\begin{figure} [t]
	\begin{center}
		\begin{tabular}{c}
			\includegraphics[width=.82\textwidth]{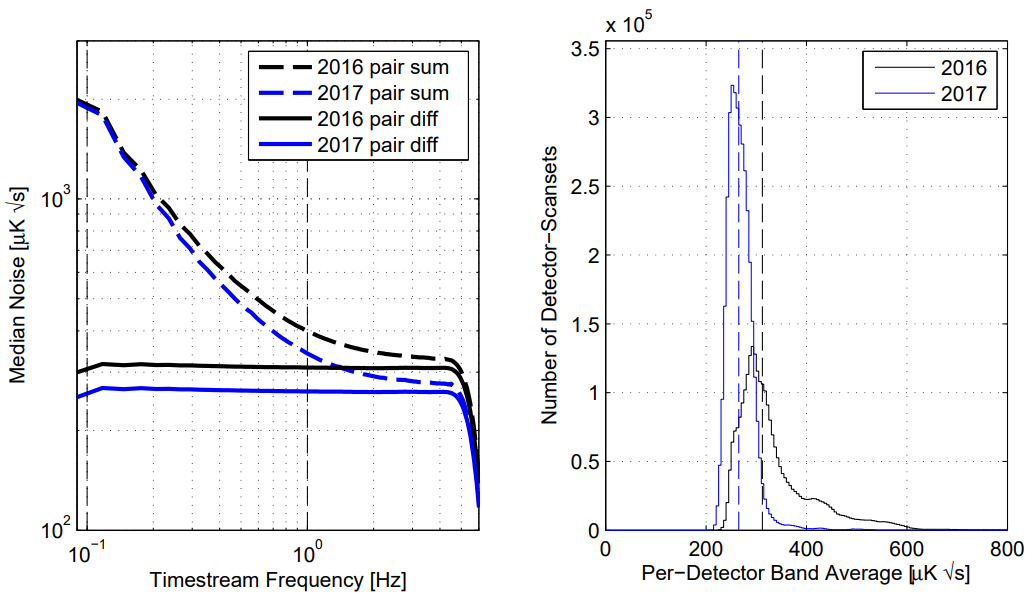}
		\end{tabular}
	\end{center}
	\caption[example] 
	{ \label{fig:NETspec} 
		\textit{Left}: Median per-detector noise spectra for \bicep3 2016 and 2017 season data, from pair-summed and pair-differenced timestreams. The pair-differenced timestreams are subjected to a third-order polynomial filtering for each half scan to reject $1/f$ noise. 
		\textit{Right}: Histogram of the per-detector per-scanset noise, applying 3rd-order polynomial timestream filtering and averaged across the 0.1-1Hz science band. Median values of 312\ukrts\, and 265\ukrts\, are marked by vertical dashed lines for the 2016 and 2017 data, respectively.}
	\end{figure} 
	
    The detector performance can be assessed by estimating NET based on timestream data. For each scanset, we take pair-summed and pair-differenced timestream data and apply data quality cuts to keep only well-behaving detector pairs. We convert them to CMB temperature by calibrating the measured temperature map to 100GHz Planck temperature map for each season. We take the power spectral density (PSD) across each azimuthal scan and multiply by two to get per-detector estimate instead of per-pair estimate. The median spectra across all detectors and scanasets are plotted on the left panel of Figure~\ref{fig:NETspec}. The pair-differenced timestreams are subjected to a third-order polynomial filtering for each half scan to reject $1/f$ noise. The noise level is reduced in the 2017 season. We find the mean across the science band $0.1-1Hz$ per detector per scanset and make the histogram over each season on the right panel of Figure~\ref{fig:NETspec}. The median value improved from $312$\ukrts\, to $265$\ukrts\, from 2016 to 2017, which is better than the $288$\ukrts\, for Keck 95GHz detectors. Reduced in-band instrumental loading and a more uniform distribution of optical efficiency contributes to the lower and tighter distribution of NET histogram in the 2017 season. The receiver NET improved from 8.68\ukrts\, to 6.64\ukrts. 
	
    \subsection{CMB maps}
    \begin{figure} [t]
		\begin{center}
			\begin{tabular}{c} 
				\includegraphics[width=0.6\textwidth]{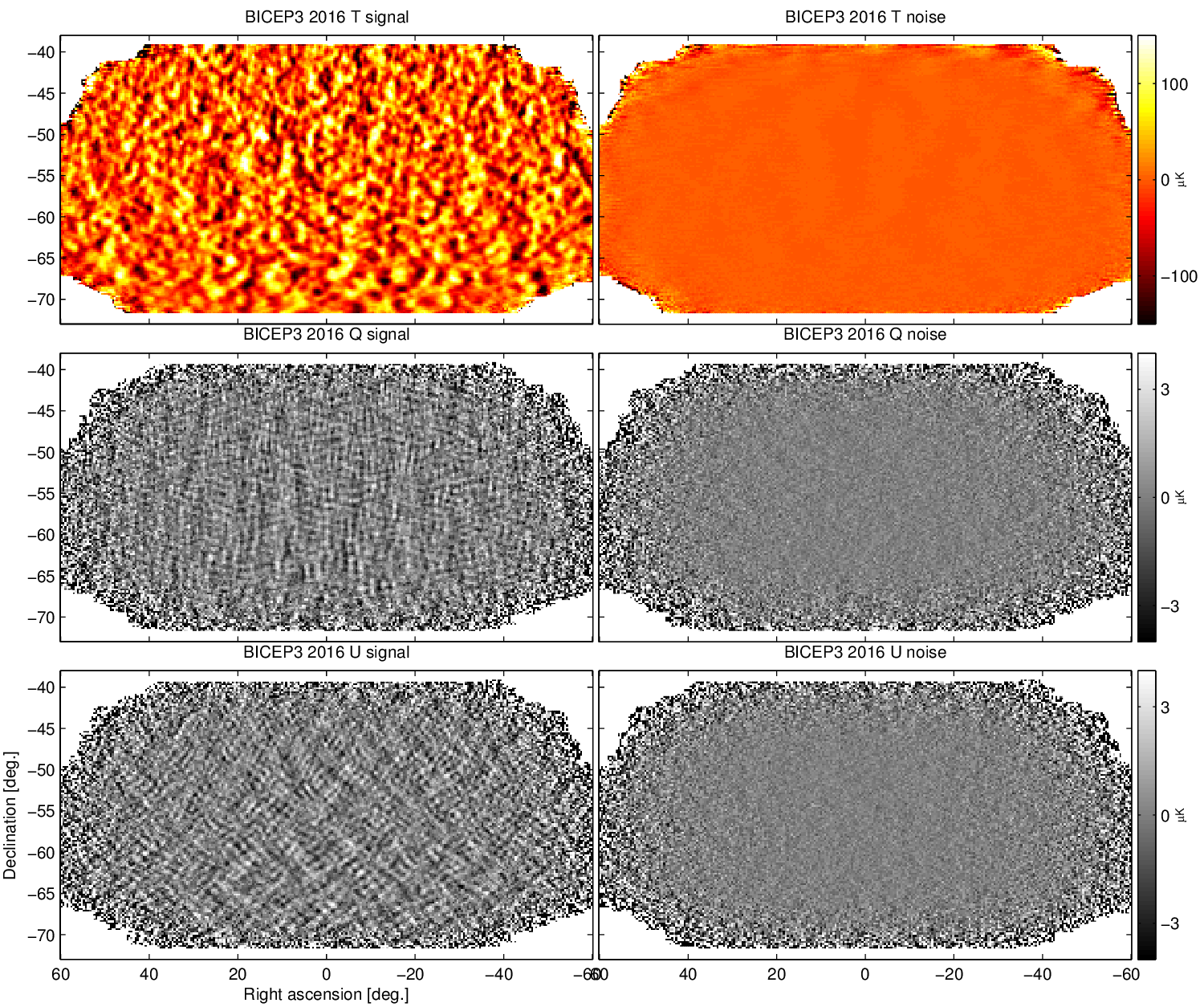}
			\end{tabular}
		\end{center}
		\caption[example] 
		{ \label{fig:B2016map} Temperature, Q and U polarization maps and their noise maps from \bicep3 2016 season.}
	\end{figure} 
	\begin{figure} [t]
		\begin{center}
			\begin{tabular}{c}
				\includegraphics[width=.82\textwidth]{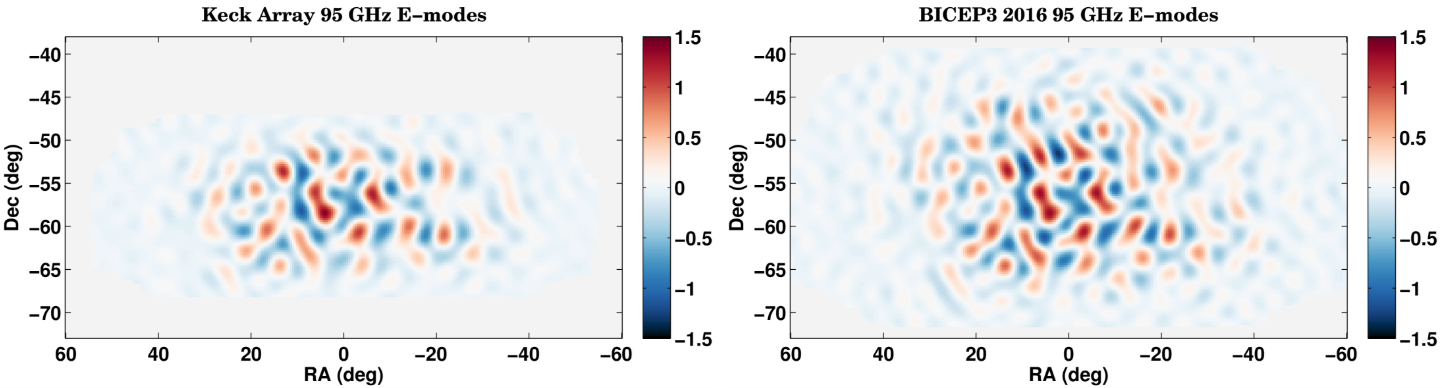}
			\end{tabular}
		\end{center}
		\caption[example] 
		{ \label{fig:K95B3Emodes} E-mode maps representing the modes within $50<\ell<120$ from Keck 95GHz receivers for two years of data (\textit{Left}) and \bicep3 95GHz receiver from 2016 season (\textit{Right}). Clearly \bicep3 observes the same E-mode sky as Keck 95GHz receivers did. In addition, with its larger field of view, \bicep3 reveals more modes.}
	\end{figure} 
     Figure~\ref{fig:B2016map} shows the temperature, Q and U polarization maps and their noise realizations from the full 2016 season data. We obtain the noise realizations by randomly flipping the sign of data from the scansets while coadding the map (details in Section V.B of Ref.~\citenum{BKI}). Figure~\ref{fig:K95B3Emodes} shows the E-mode maps from Keck 95GHz receivers with two years of observation and from \bicep3 with 2016 season data. The maps of the Stokes parameters Q and U were Fourier-transformed, converted to E- and B-modes, taken with only the modes within $50<\ell<120$, and Fourier-transformed back to the maps. Figure~\ref{fig:K95B3Emodes} clearly shows the agreement between the two experiments as well as additional modes \bicep3 observes in the larger field.
	
	\subsection{Preliminary 2018 Season Performance}
	\begin{figure} [t]
	\begin{center}
		\begin{tabular}{c} 
			\includegraphics[width=0.4\textwidth]{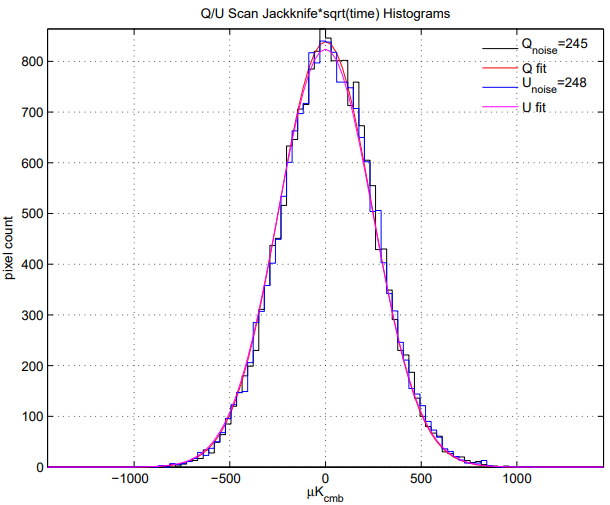}
		\end{tabular}
	\end{center}
	\caption[example] 
	{ \label{fig:MapNET18} Histogram of the integration-time-weighted noise map pixels for the map-based NET estimate for the 2018 preliminary data set.}
	\end{figure} 
	During the 2017-18 austral summer, we kept the \bicep3 receiver as it was in 2017 season. We present the preliminary performance of 2018 season using the early $\sim$406 hours of CMB data (from 719 scansets)\footnote{Each scanset spans about 50 minutes, but the time \bicep3 spends on the CMB source is about 34 minutes after removing the time for calibration and turning around the scanning direction.} and confirm that \bicep3 achieves similar performance in 2017 and 2018.
	
	The timestream-based NET estimate agrees with the results of the 2017 season. When we scale the histogram of the per-detector per-scanset noise as in Figure~\ref{fig:NETspec} for 2018 preliminary data to the number of observed scansets in 2017, the histograms fit together. The median timestream-based per-detector NET is 262\ukrts\, and the receiver NET is 6.56\ukrts.
		
	Map based NET is another estimate of noise performance based on the standard deviation across all map pixels of a polarized noise map while accounting for weighted integration time of each pixel. We split the data into two halves according to the two azimuthal scan directions. By differencing the two halves, we remove the signal and obtain noise polarization maps. The noise level at each pixel, $n_p$, is related to the NET at that pixel by $n_p=NET_p/\sqrt{t_p}$ where $t_p$ is the integration time at pixel $p$. We multiply the noise map with the map of the square root of the integration time. Then we form a histogram for all the pixels that are within $\sqrt{2}$ standard deviations from median pixel values of the polarization variance maps to exclude noisy edge pixels, although they get down-weighted by short integration time. We fit a Gaussian to the histogram, and the standard deviation is taken as the estimate of the NET. The histogram (Figure~\ref{fig:MapNET18}) for the 2018 season preliminary data set exhibits good performance, achieving per-detector NET of 248\ukrts\, and receiver NET of 6.62\ukrts. This estimate is slightly different from the timestream-based estimation as it assumes Gaussian noise distribution and considers the full range of spectra instead of averaging across 0.1-1 Hz. The receiver NET is evaluated by dividing the estimated per-detector NET by the effective count of detectors, which is estimated by dividing the total polarization integration time of all detectors by the total wall-clock observing time.

	\section{CONCLUSION}
	In this proceeding, we present the upgrades we performed on \bicep3 during the 2016-17 austral summer. We have evaluated the noise performance of the detectors from the full season data in 2016 and 2017 as well as the preliminary performance from 2018 season. We confirm the improved performance in 2017 continues to hold in 2018.
	
	The major upgrades include the replacement of the IR filter stack formed of reflective metal-mesh filters with absorptive foam filters, replacing delaminated edge-filters and replacing four tile modules. Comparing the full season data in 2016 and 2017, we show improvement in the timestream-based median per-detector NET from 312\ukrts\, to 265\ukrts\, and the array NET from 8.68\ukrts\, to 6.64\ukrts. \bicep3 performs better than Ref.~\citenum{wu15}'s projected receiver NET of 6.9\ukrts. \bicep3 continues to make the deepest CMB polarization map made to date at 95 GHz.
        
    \bicep3 still has room for improvement. Ref.\citenum{Barkats18} suggests the usage of ultra-thin window design would further reduce the in-band loss and may improve NET. The change from \bicep3's 1.25-inch thick window to 0.01-inch thick window may reduce NET by 20\% at 95GHz. \bicep3 has demonstrated stable performance and serves as a stepping stone for the upgrade of the Keck Array to the \bicep Array, an array of four \bicep3-size receivers observing at 30/40, 95, 150, and 220/270 GHz\cite{Grayson16,Hui18}. Research and development is ongoing, including the window design\cite{Barkats18}, cryostat and mount design\cite{Crumrine18}, module design\cite{Soliman18}, and readout technology\cite{Henderson18,obrient2018tkid}, which are included in this volume of proceedings.

	\acknowledgments 
	The \bicep/Keck projects have been made possible through a series of grants from the National Science Foun-
dation including 0742818, 0742592, 1044978, 1110087, 1145172, 1145143, 1145248, 1639040, 1638957, 1638978,
1638970, \& 1726917, by the Gordon and Betty Moore Foundation, by the Keck Foundation, and by the grant 55802 from John Templeton Foundation. The development of antenna-coupled detector technology was supported by the JPL Research and Technology Development Fund and NASA Grants 06-ARPA206-0040, 10-SAT10-0017, 12-SAT12-0031, 14-SAT14-0009 \& 16-SAT16-0002. The development and testing of focal planes were supported by the Gordon and Betty Moore Foundation at Caltech. Readout electronics were supported by a Canada Foundation for Innovation grant to UBC. The
computations in this paper were run on the Odyssey cluster supported by the FAS Science Division Research
Computing Group at Harvard University. The analysis effort at Stanford and SLAC is partially supported by
the U.S. DoE Office of Science. We thank the staff of the U.S. Antarctic Program and in particular the South
Pole Station without whose help this research would not have been possible. We thank all those who have contributed past efforts to the \bicep/Keck series of experiments, including the \bicep1 team. Tireless administrative support was provided by Kathy Deniston, Sheri Stoll,
Irene Coyle, Donna Hernandez, Dana Volponi, and Julie Shih.
	
	\bibliography{report} 
	\bibliographystyle{spiebib} 
	
\end{document}